\documentclass[runningheads]{llncs}

\usepackage{graphicx}

\begin{document}

\title{Requirements-Driven Visualizations for Big Data Analytics: a Model-Driven approach}

\titlerunning{Model Driven Visual Analytics}

\author{Ana Lavalle\inst{1}\and
Alejandro Mat{\'e}\inst{1}\and
Juan Trujillo\inst{1}}

\authorrunning{A. Lavalle et al.}

\institute{Lucentia (DLSI), University of Alicante, Carretera San Vicente del Raspeig s/n, 03690, San Vicente del Raspeig, Alicante, Spain \\ Lucentia Lab, C/ Pintor Pérez Gil, N-16, Alicante \\ \email{\{alavalle,amate,jtrujillo\}@dlsi.ua.es}}

\maketitle    

\begin{abstract}
Choosing the right Visualization techniques is critical in Big Data Analytics. However, decision makers are not experts on visualization and they face up with enormous difficulties in doing so. There are currently many different (i) Big Data sources and also (ii) many different visual analytics to be chosen. Every visualization technique is not valid for every Big Data source and is not adequate for every context. In order to tackle this problem, we propose an approach, based on the Model Driven Architecture (MDA) to facilitate the selection of the right visual analytics to non-expert users. The approach is based on three different models: (i) a requirements model based on goal-oriented modeling for representing information requirements, (ii) a data representation model for representing data which will be connected to visualizations and, (iii) a visualization model for representing visualization details regardless of their implementation technology. Together with these models, a set of transformations allow us to semi-automatically obtain the corresponding implementation avoiding the intervention of the non-expert users. In this way, the great advantage of our proposal is that users no longer need to focus on the characteristics of the visualization, but rather, they focus on their information requirements and obtain the visualization that is better suited for their needs. We show the applicability of our proposal through a case study focused on a tax collection organization from a real project developed by the Spin-off company Lucentia Lab.

\end{abstract}
\section{Introduction}

Data is continuously growing, specially since the last decade. With ever larger amounts of data that need to be interpreted and analyzed, using the right visualizations is crucial to help decision makers to properly analyze the data and guide them to take better informed decisions. 

In this new era of Big Data Analytics, there has been an increasing interest from both the academic and industry worlds in different phases of the data life cycle: from the storage to the analysis, cleaning or integration and, of course, the visualization. Data and Information Visualization are becoming strategic for the exploration and explanation of large data sets due to the great impact that data have from a human perspective. An effective, efficient and intuitive representation of the analyzed data may result as important as the analytic process itself \cite{komis17}. However, larger data sets and their complexity in terms of heterogeneity contribute to make the representation of data more complex \cite{caldarola2016improving}. 

In this context, defining and implementing the right visualization for a given data set is a complex task for companies, specially in the age of Big Data where heterogeneous and external data sources require knowledge of the underlying data to create an adequate visualization. As such, choosing the wrong visualizations and misunderstanding the data leads to wrong decisions and considerable losses. One of the key difficulties for defining the right visualization technique is the lack of expertise in information visualization of decision makers. Another critical aspect is that, apparently, a large set of visualizations may \textbf{}be equally valid for any given data sets, which has been proven to be absolutely wrong \cite{vartak2015s}, each data set and each analysis has its particular characteristics and not always all the types of visualization are valid to represent them.

\setlength{\belowcaptionskip}{-12pt}

In order to tackle the above-presented problems, we propose an approach, based on the Model Driven Architecture (MDA) \cite{MDA2.0} proposed by the Object Management Group (OMG).

\begin{figure*}[bp!]
\centering
\includegraphics[width=0.96 \textwidth]{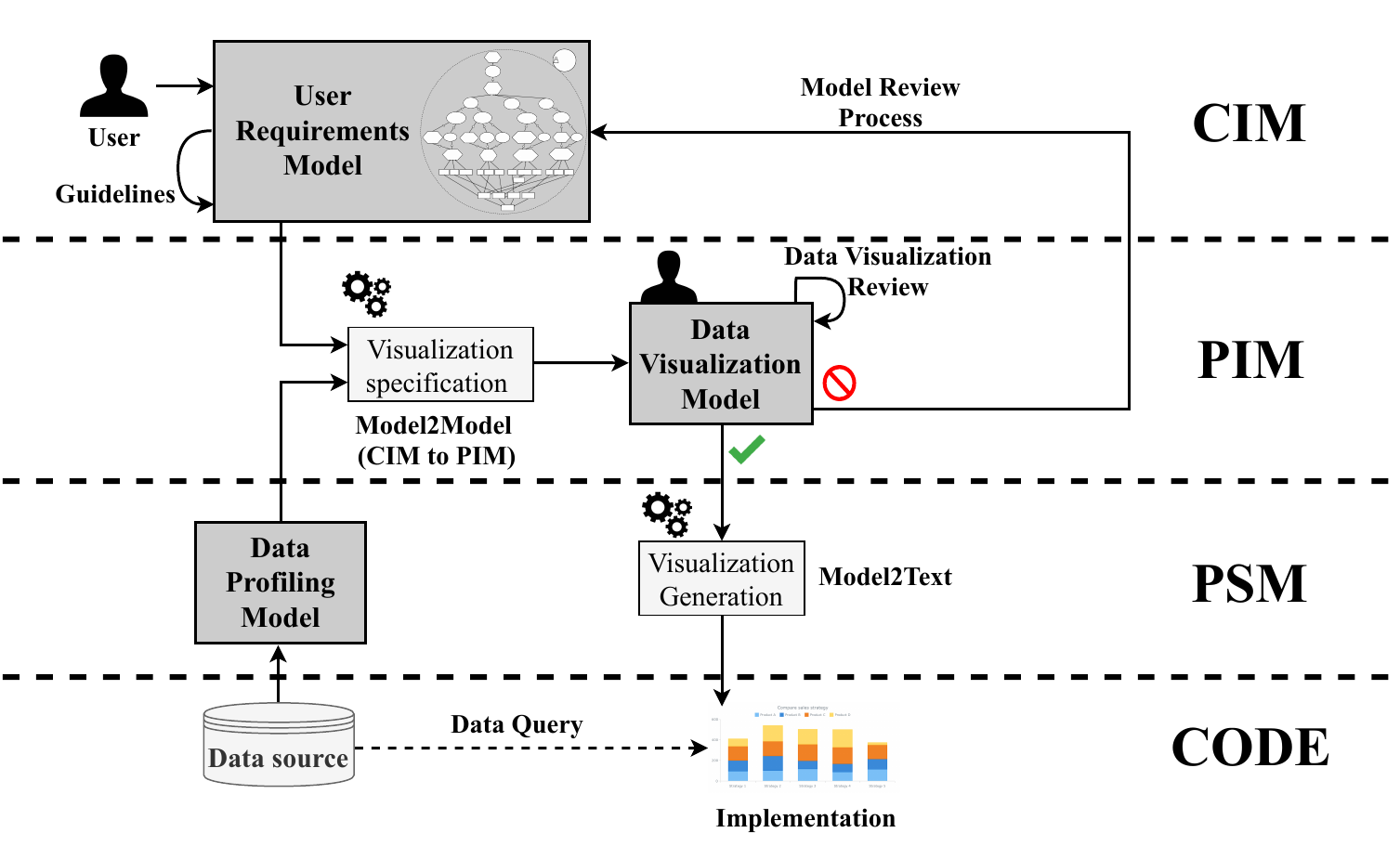}
\caption{Overall view of the process proposed.}
\label{fig:process}
\end{figure*}

Fig. \ref{fig:process} summarizes the process followed in our proposal, aligned with MDA. Firstly, a sequence of questions guides users in creating a Goal-Oriented \cite{mate2014adding} model that captures their needs. This model (CIM layer) enables them to capture all the visualizations that are needed to tackle their information needs. The user requirements together with the data profiling information coming through the Data Profiling Model (at PSM layer) are used as a visualization specification that is input to a model to model transformation. This transformation generates the Data Visualization Model (PIM layer). This model allows users to specify exactly how they need to visualize the data. It also allows them to determine if the proposed visualization is adequate to satisfy the essential requirements for which it was created. The validation process is performed through a questionnaire according to user goals model. If the proposed visualization passes the validation. Otherwise, an unsuccessful validation points out to the existence of missing or wrongly defined requirements that must be reviewed. This process is repeated until all user requirements are fulfilled. Finally, a model to text transformation generates the implementation of the visualization using the data visualization model as input.

The great advantage of our proposal is that users no longer focus on the underlying technical aspects or finding the most adequate visualization technique to be used in every different data analytic process. By following our approach, decision makers obtain the visualization technique that is better suited to their information needs and the characteristics of the data at hand in a semi-automatic way. This is achieved thanks to our alignment with MDA, enabling us to incrementally refine the visualization until its implementation is obtained. 

The rest of the paper is structured as follows. Section 2 presents the related work in this area. Section 3 presents the different proposed models of the approach based on the MDA. Section 4 discusses a real case study in the fiscal domain. Finally, Section 5 summarizes the conclusions and our future work.

\section{Related Work}

Several works have focused on proposing different ways to find the best visualization. \cite{borner2014atlas} surveys the main classifications proposed in the literature and integrates them into a single framework based on six visualization requirements. In \cite{madhuiba}, authors propose a framework for choosing the best visualization where the main types of charts are related to users goals and to data dimensionality, cardinality, and data type they support. Finally, \cite{golfarelli2019goal} proposes a model to automate the translation of visualization objectives specified by the user into a suitable visualization type based on seven visualization requirements.

Additionally, several approaches are focused on the analysis of visualization representations. \cite{oliveira2017meta} describes an information visualization taxonomy. \cite{pena2017big} make a revision of visualization techniques for Big Data to determine which are the most optimistic when analyzing Big Data. \cite{bull2005visualization} propose a metamodel to represent tree and graph views by modeling nodes and edges. Similarly, \cite{domokos2002open} uses nodes and edges to draw basic shapes like lines and circles. 

Other works are focus on visual analytics recommendation systems. \cite{vartak2017towards} detail the key requirements and design considerations for a visualization recommendation system and identify a number of challenges in realizing this vision and describe some approaches to address them. \cite{du2016eventaction} propose EventAction, a prescriptive analytics interface designed to present and explain recommendations of temporal event sequences. Additionally, \cite{vartak2015s} propose SEEDB, a visualization recommendation engine to facilitate fast visual analysis, SEEDB explores the space of visualizations, evaluates promising visualizations for trends, and recommends those it deems most “useful” or “interesting”. In \cite{morgan2017vizdsl} authors propose a new language VizDSL for creating interactive visualizations that facilitate the understanding of complex data and information structures for enterprise systems interoperability.

To the best of our knowledge, the only approaches that follow the MDA philosophy in the Big Data Context are presented within the TOREADOR project. In \cite{ardagna2017model}, the authors propose a Model-driven approach that aims to lower the amount of competences needed in the management of a Big Data pipeline. 
\cite{leida2016facing} illustrates a use case exploiting the Model-driven capabilities of the TOREADOR platform as a way to fast track the uptake of business-driven Big Data models. \cite{misale2017comparison} provides a layered model that represents tools and applications following the Dataflow paradigm.

Despite all the work presented so far, none of the approaches provide a way to easily translate user requirements into visual analytics implementations. Furthermore, there is an absence of a methodology that guides users in obtaining the most adequate visualization, allowing them to focus on their own needs rather than on the characteristics of the visualization.

\section{A MDA approach for Visual Analytics}

As previously introduced in the paper, specifying the right visualization for a user is a challenging task. User has not only to take into account her needs, which are on a completely different abstraction level, but also consider characteristics of the data that make inadequate the use of certain visualizations. In order to let the user focus on her information needs, we aim to bridge the gap between the user requirements and their visualization implementation.

To this aim, we propose a development approach Fig. \ref{fig:process} in the context of the Model Driven Architecture \cite{MDA2.0}. Our main goal is to help users to generate the visualizations that are better suited to meet their information needs. Following the basic principles of MDA, our proposal builds on three types of models:

\begin{itemize}
\item \textbf{User Requirements Model (CIM layer):} Allows users to capture their information needs and certain visualization aspects that are needed to tackle them.

\item \textbf{Data Visualization Model (PIM layer):} Enables users to specify the characteristics of their visualizations before obtaining their implementation.

\item \textbf{Data Profiling Model (PSM layer):} Abstracts the required information from the data sources to i) aid in determining the most adequate visualization and ii) take certain aspects of data into account for their representation (such as whether they are numeric or categorical).  
\end{itemize}

The process starts by capturing information needs at the CIM level. Then, a data profiling process is run to generate a data profiling model at the PSM level that contains the relevant data characteristics for the process. Once both models have been obtained, they are processed through a model to model transformation that generates a data visualization model at the PIM level. This model provides the user with the better suited visualization for her needs and the data available, and allows her to modify different aspects of the visualization such as the axis where each attribute should be positioned, the orientation, or the color range among others. Once the model refinement process is finished, a model to text transformation generates the implementation using a visualization library, such as D3.js in our case.

\subsection{User Requirements Model}

Our approach starts from a goal-oriented requirements model that allows us to capture information needs. To describe the coordinates required to build a visualization context (\textit{Goal, Interaction, User, Dimensionality, Cardinality, Independent Type, and Dependent Type}) we follows the specification to automate data visualization in Big Data Analytics given in \cite{golfarelli2019goal}, in this way we make sure that the visualization specification is addressed in terms of Big Data. Due to paper constraints, we cover only the main aspects of our requirements model.

\begin{figure*}[bp!]
\centering
\includegraphics[width=1\textwidth]{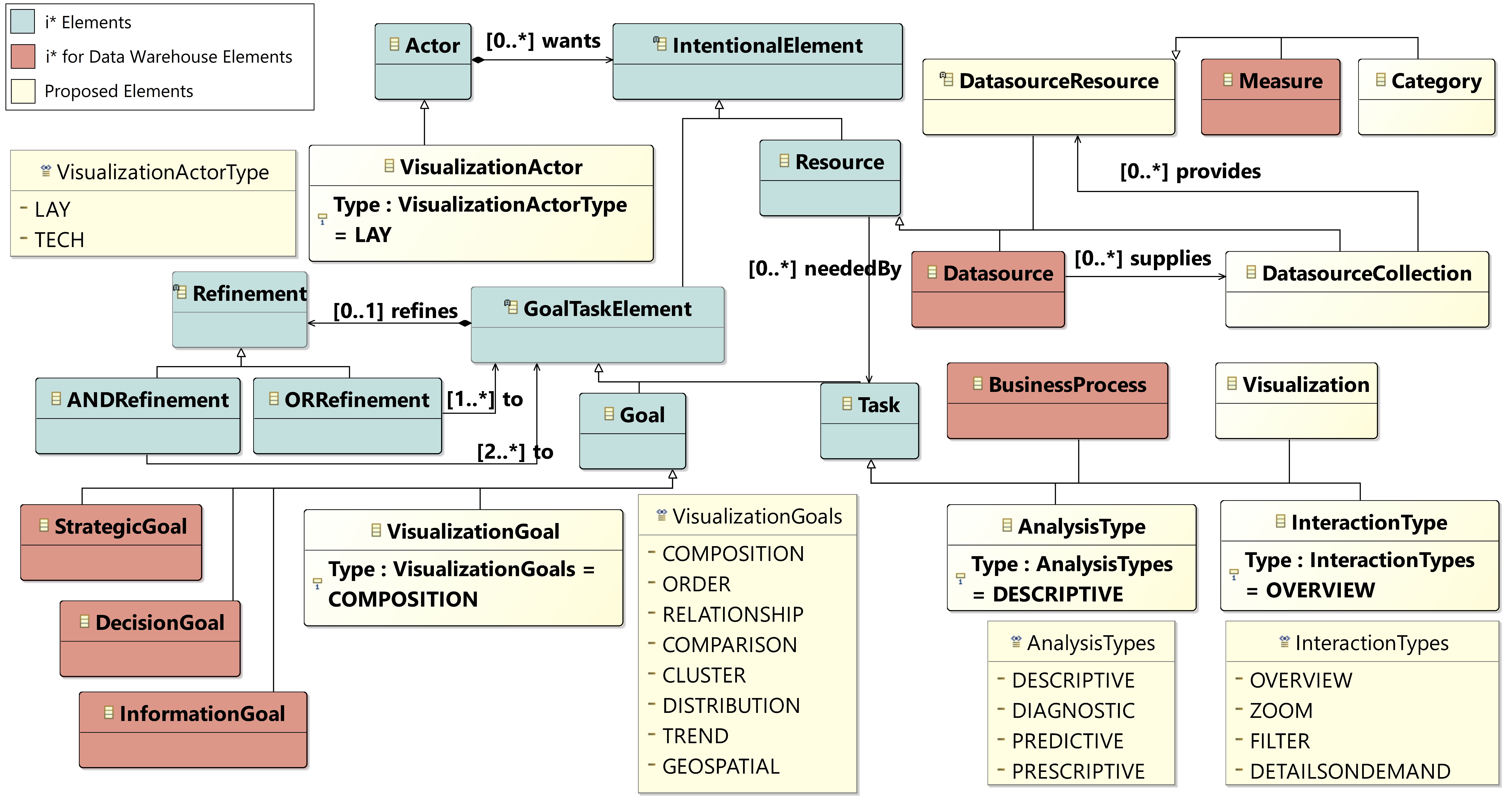}
\caption{User Requirements Metamodel.}
\label{fig:metamodel}
\end{figure*}

Our metamodel shown in Fig. \ref{fig:metamodel} is an extension of i* and the \emph{i* for Data Warehouses} extension \cite{mate2014adding}. Existing elements in the i* core are represented in blue (light grey), whereas those in i* for Data Warehouses are represented in red (dark grey). The new concepts added by our proposal are represented in white. 

The first element is the \textit{VisualizationActor}, which models the user of the system. There are two types of Visualization Actors: Lay, if she has no knowledge of complex visualizations, and Tech, if she has previous experience and is accustomed to Big Data Analytics. 
Next is the \textit{BusinessProcess} on which users will focus their analysis. The business process will serve as the guideline for the definition of \textit{Goals}.

The \textit{AnalysisType} allows users to express which kind of analysis they wish to perform. The type of analysis can be determined by selecting which question from the following ones \cite{shi2017data} is to be answered:
How to act? (Prescriptive), Why has it happened? (Diagnostic), What is going to happen? (Predictive) or What to do to make it happen? (Descriptive).

Next, a Visualization represents a specific visualization that will be implemented to satisfy one or more \textit{VisualizationGoals}. Each \textit{VisualizationGoal} describes an aspect of the data that the visualization should reflect. These goals can be Composition, Order, Relationship, Comparison, Cluster, Distribution, Trend, or Geospatial, as considered in \cite{golfarelli2019goal}.

Along with \textit{VisualizationGoals}, Visualizations have one or more \textit{InteractionTypes}, that capture how the user will interact with the visualization. The different kinds of interaction are Overview, Zoom, Filter, or Details on Demand as \cite{golfarelli2019goal} consider to data visualization in Big Data Analytics. Finally, a Visualization makes use of one or more \textit{DatasourceResource} elements which feed the data to the visualization. 

Using these concepts we allow users to define their needs instead of focusing on technical details that are not relevant at this level.

\subsection{Data Profiling Model}
Our second model is the Data Profiling Model. This model captures the data characteristics that are relevant to the visualization and is generated through a data profiling process. Firstly, users will select the data sources that they want to be represented in the visualization. Consecutively, the data analyst will analyze the data sources extract the values of the coordinates by analyzing the features of the data sources. In this way, users do not need to manually inspect the data or have a deep understanding.

To know how to delimit the values for each coordinate we have use the values proposed in \cite{golfarelli2019goal} to Big Data Analytics. In this way we classify the Dimensionality, Cardinality, and Dependent/Independent Type as follows:

\textbf{Cardinality} represents the cardinality of the data. It can either be \textit{Low} or \textit{High}, depending of the numbers of items to represent. \textit{Low} cardinality considers a few items to a few dozens of items while \textit{High} cardinality is set if there are some dozens of items or more.

\textbf{Dimensionality} is used to declare the number of variables to be visualized. Specifically, it can be \textit{1-dimensional} when the data to represent is a single numerical value or string, \textit{2-dimensional} when one variable depends on other, \textit{n-dimensional} when a data object is a point in an n-dimensional space, \textit{Tree} when a collection of items have a link to one other parent item, or \textit{Graph} when a collection of items are linked to arbitrary number of other items.

\textbf{Type of Data:} is used to declare the type of each variable. It can be \textit{Nominal} when each variable is assigned to one category, \textit{Ordinal} when it is qualitative and categories can be sorted, \textit{Interval} when it is quantitative and equality of intervals can be determined, or \textit{Ratio} when it is quantitative with a unique and non-arbitrary zero point.

\subsection{Visualization specification Transformation - (Model to Model)}

\begin{figure*}[bp!]
\centering
\includegraphics[width=0.78 \textwidth]{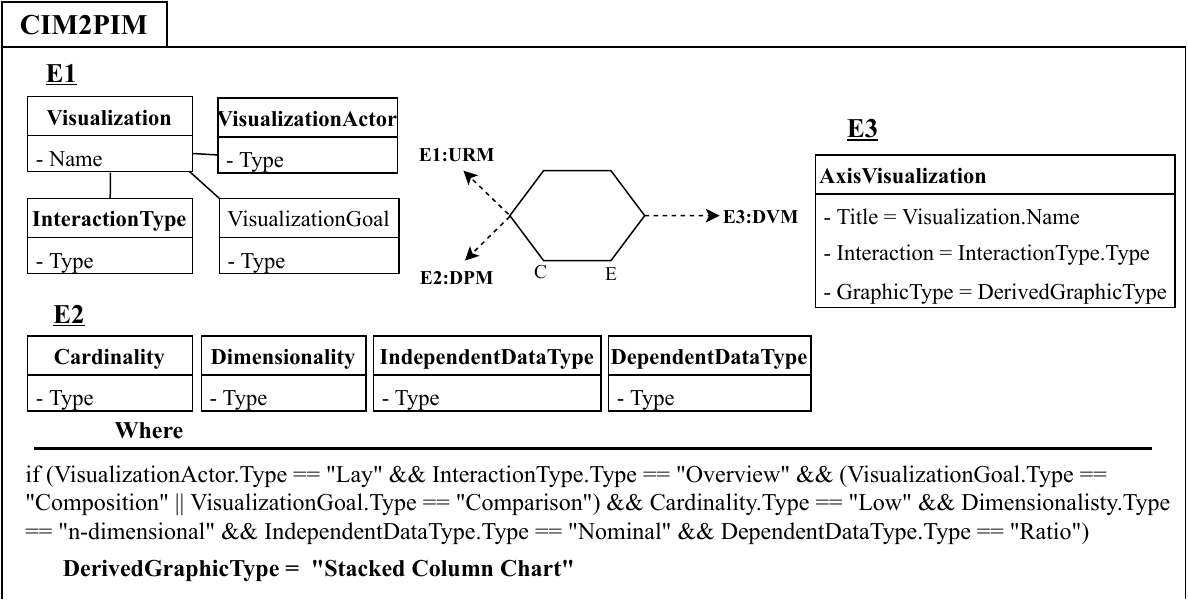}
\caption{Generation of axis based visualizations from user requirements.}
\label{fig:trans1}
\end{figure*}

\begin{figure*}[bp!]
\centering
\includegraphics[width=0.78 \textwidth]{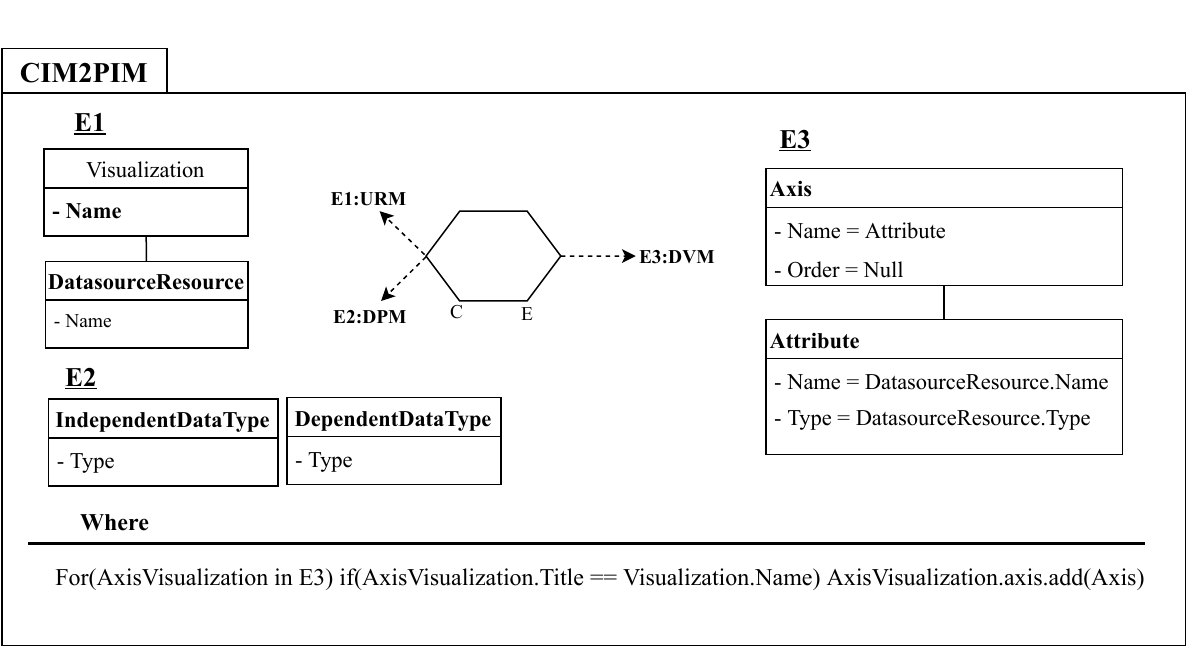}
\caption{Generation of axes for axis based visualizations from user requirements.}
\label{fig:trans2}
\end{figure*}

Information coming from User Requirements Model and the Data Profiling Model form the Visualization Specification. This specification is transformed into a data visualization model using a set of model to model transformations, presented in Fig. \ref{fig:trans1} and Fig. \ref{fig:trans2} by the OMG standard language QVT \cite{QVT}. According to the nature of the visualization to be derived, there are two types of transformations. On the one hand, we can have axis-based visualizations, such as column chart, line chart, bubble chart, etc. On the other hand, some visualizations such as dendogram, chord or graphs require graph-based visualizations, which make use of nodes and edges instead of axis.

Due to space constraints, we will focus on how axis-based visualizations are derived. Our first transformation (Fig. \ref{fig:trans1}), generates the visualization element, an AxisVisualization in this case. An AxisVisualization is derived according to the graphic type established by the transformation. This value is derived using the imperative part of the transformation (Where clause) according to the specific criteria established by \cite{golfarelli2019goal} for the each graphic type. The values Cardinality, Dimensionality, IndependentDataType and DependentDataType are obtained from the data profiling. Finally, the visualization name and interaction type defined in the User Requirements Model are used to establish the title and interaction of the Axisvisualization. 

Next, as Fig. \ref{fig:trans2} shows, each of the axes is generated individually. An axis is generated for each measure or category (abstracted by the DatasourceResource element) in the User Requirements Model. Afterwards, each axis is assigned their corresponding visualization by iterating over the data visualization model, completing the derivation of the visualization.

\subsection{Data Visualization Model}

\begin{figure*}[bp!]
\centering
\includegraphics[width=0.8 \textwidth]{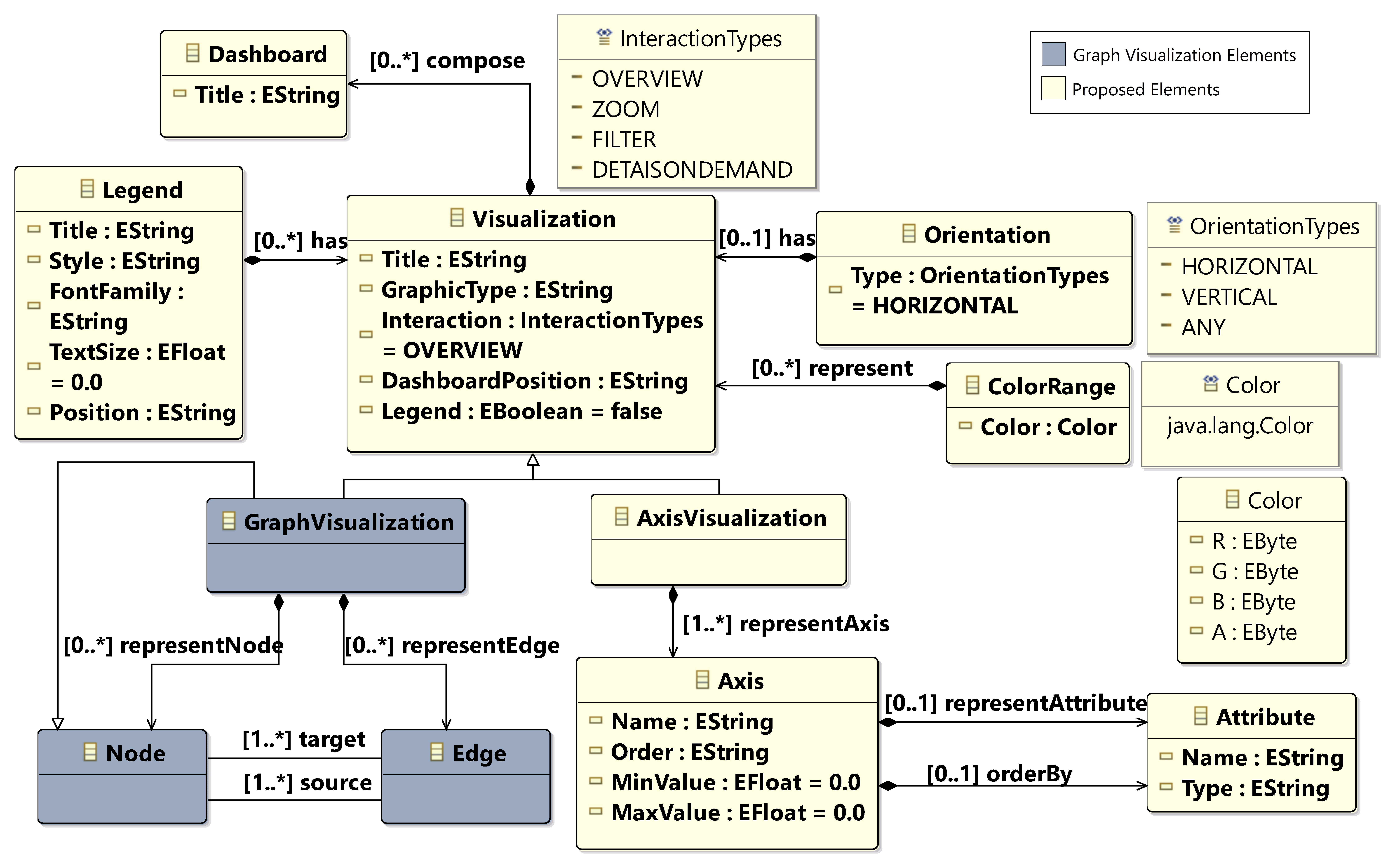}
\caption{Data Visualization Metamodel.}
\label{fig:metamodelVis}
\end{figure*}

In order to verify if the recommended visualization is adequate to satisfy the information needs of the user and allow her to customize each visualization, we require an abstraction of the visualization to be generated. Despite our best efforts, there is no metamodel proposed so far to model visual analytics. Thus, to support our process, we have defined a novel visualization metamodel. 

Our metamodel shown in Fig. \ref{fig:metamodelVis}. is composed of elements extracted from \cite{bull2005visualization} to define tree and graph visualizations, represented in blue (grey) color, while new concepts added by our proposal to detail the specification of visualizations represented in white. In the following, we describe the concepts included in the proposed metamodel.

The main element is \textit{Visualization}, this element collects all the visualization requirements that should be met. It contains a visualization Title; a Legend, that may be shown or not; a Graphic Type that determines the type of visualization; a set of interactions that contain the type of interaction that must be supported (Overview, Zoom, Filter or Detais-on-demand); and a Dashboard Position, in the event that the visualization will be part of a \textit{Dashboard}. 

In order to define the representation of a visualization, other elements are necessary. A visualization has and \textit{Orientation}, either Horizontal, Vertical, or Any (when the graphic type does not have orientation). Moreover, a visualization has a \textit{ColorRange}, that represents the range of colours that will be used by the visualization, an aspect of special importance for color-blind users.

A visualization will be instanced as either a \textit{GraphVisualization} or an \textit{AxisVisualization} depending on the type of visualization. A \textit{GraphVisualization} may contain several \textit{Nodes} and \textit{Edges} \cite{bull2005visualization}. Meanwhile, an \textit{AxisVisualization} constains a series of axes that represent the data. An \textit{Axis} is may have a Name, Order, Minimum Value and Maximum Value. Each Axis represents an Attribute at most. An \textit{Attribute} has a Name and a Type. Attributes can be used to be represented or to set the order of the data in the visualization.

\subsection{Visualization Generation Transformation - (Model to Text)}

The Visualization Generation Transformation has as input the data visualization model from the previous step. This transformation transforms each element within the visualization specification into a code level specification for a graphic library. In our case, we use the D3 JavaScript library \cite{D3js} for generating the visualization. The GraphicType and the Orientation determine the type of visualization to implement. Categories and measures and their respective axes  determine how the data is assigned to each axis. Meanwhile, the Color Range is translated into custom color scales. Moreover, if a Legend has been defined, the type of the legend, title, position, font family and text size are be translated attributes in the corresponding d3.legend function call. Finally, the title is used to provide a name to the visualization created, and the dashboard position is used to assign a position to the visualization.

\section{Case Study}

In order to evaluate the validity of our approach we have applied it to a real case study, based on a tax collection organization. Due to space constraints, we provide a reduced example including enough data in order to allow readers to completely understand the approach. Therefore, the example is constrained to a Tax Region Area covering only three provinces. The organization requires a set of visualizations to analyze their data in order to help them detect underlying patterns in their unpaid bills and tax collection distribution. Due to the sensitivity of their data, we are not allowed to show the real values.

\subsection{Specifying User Requirements}

\begin{figure*}[tp!]
\centering
\includegraphics[width=0.93 \textwidth]{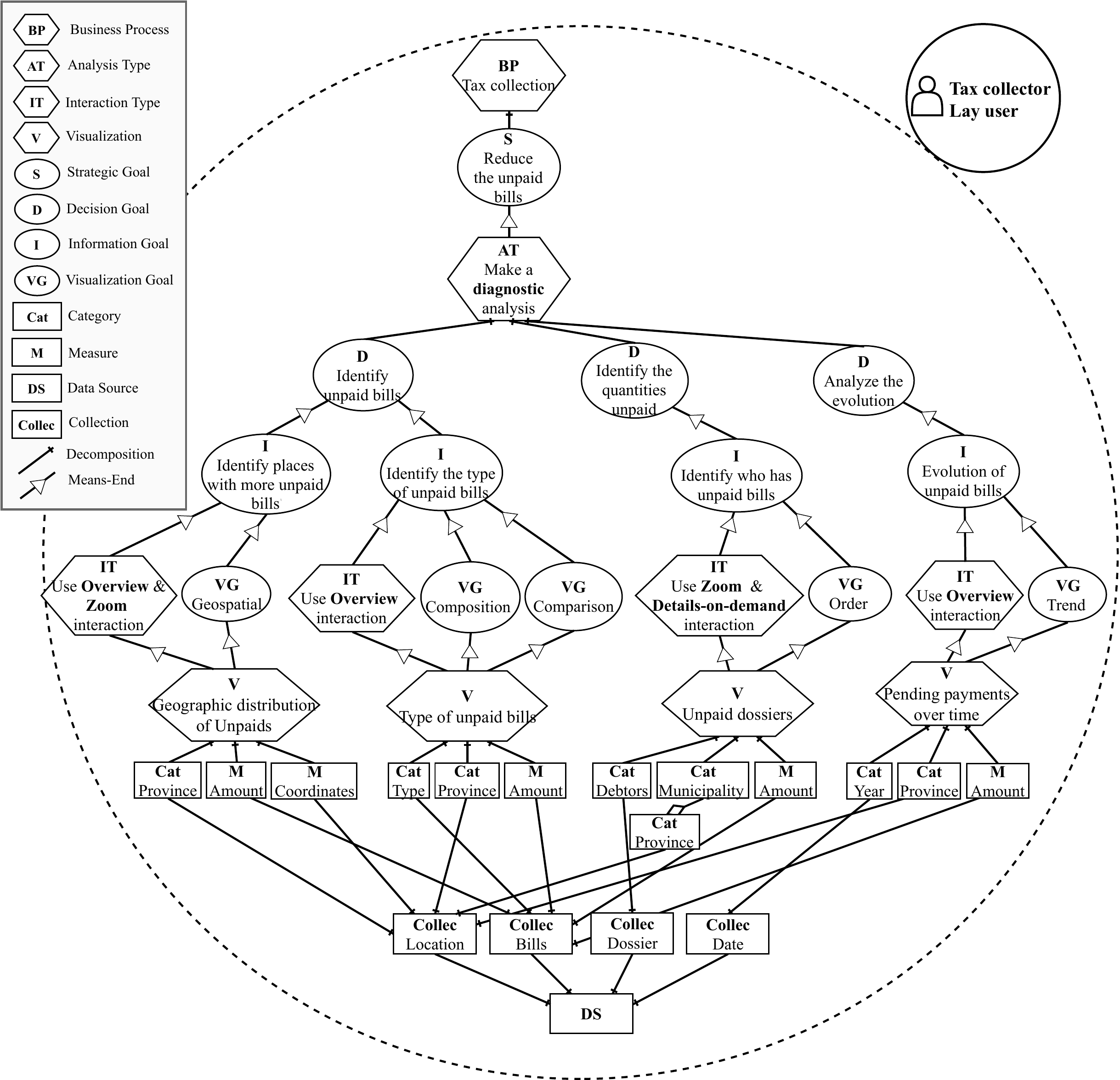}
\caption{Application of our user requirements metamodel to the case study.}
\label{fig:example}
\end{figure*}

Through the application of our User Requirements Model to a tax collection organization, the Fig. \ref{fig:example} has been generated. A tax collector user wants to analyze the unpaid debts. Therefore, the analysis will focus on the \textit{``Tax collection''} business process. Defining a business process helps determining the scope of the analysis and the goals pursued. The user is not a specialist in Big Data Analytics but rather an expert in tax management, thus she is defined as \textit{``Lay user''}.

Next, the main objectives of the business process are defined as shown in Fig. \ref{fig:example}. Specifically, the user defined her strategic goal as \textit{``Reduce the unpaid bills''}. Strategic goals are achieved by means of analyses that support the decision-making process. The analysis type allows users to express what kind of analysis they wish to perform. In this case, the user wishes to know why bills are unpaid. Thus, the user decides to perform a \textit{``Diagnostic analysis''}.

The diagnostic analysis is decomposed into decision goals. The user defined her decisions goals as: \textit{``Identify unpaid bills''}, \textit{``Identify the quantities unpaid''}, and \textit{``Analyze the evolution''}. Decisions goals communicate the rationale followed by the decision-making process; however, by themselves they do not provide the necessary details about the data to be visualized. Therefore, for each decision goal we specify one or more information goals.

From each of the decision goals the user refined the following information goals: \textit{``Identify places with more unpaid bills''}, \textit{``Identify the type of unpaid bills''}, \textit{``Identify who has unpaid bills''}, and \textit{``Evolution of unpaid bills''}. Information goals represent the lowest level of goal abstraction. And for each information goal, we will have one visualization to achieve it. A visualization is characterized by one or more visualization goals which describe what aspects of the data the visualization is trying to reflect, and one or more kinds of interaction that they will like to have with the visualization. Moreover, a visualization will make use of one or more data source elements to get the relevant data from the database. In this case, the user defines the interactions she want to have with each visualization and her visualization goals following user guidelines. \textit{``Overview''}, \textit{``Zoom''} and \textit{``Details-on-demand''} have been defined as interactions and  \textit{``Geospatial''}, \textit{``Composition''} ,\textit{``Comparison''}, \textit{``Order''}, and \textit{``Trend''} as visualization goals. Finally, the user specifies the data source where the analysis will be performed and selects the Categories and Measures that will populate the visualizations.

\subsection{Profiling Data Sources}

Once user have defined the data sources and collections from where the data will be extracted, it is possible to profile data sources to determine Dimensionality, Cardinality and Dependent/Independent Type.

We focus on the \textbf{``Identify the type of unpaid bills''} Information Goal from our Goal-Oriented model, which requires information about categories \textit{``Type''} and \textit{``Province''} and measures \textit{``Amount''}. Firstly, by the Data Profiling Model, are classified the independent variables \textit{``Type''} and \textit{``Province''} as \textbf{Nominal} and the dependent variable \textit{``Amount''} as \textbf{Ratio}. Dimensionality is set to \textbf{n-dimensional}, because the user has defined 3 variables to visualize. Finally, the Cardinality is defined as \textbf{Low} Cardinality because the data contains a few items to represent 3 provinces to represent and there are 6 types of bills. 

Overall, the visualization specification obtained through User Requirements Model and Data Profiling Model are: 

\begin{itemize}
\item \textbf{Visualization Goal}: Composition \& Comparison
\item \textbf{Interaction}: Overview
\item \textbf{User}: Lay
\item \textbf{Dimensionality}: n-dimensional
\item \textbf{Cardinality}: Low
\item \textbf{Independent Type}: Nominal
\item \textbf{Dependent Type}: Ratio
\end{itemize}

With the definition of this visualization specification, by applying our visualization specification transformation, the visualization type generated is "Stacked Column Chart".

\subsection{Specifying Data Visualizations Requirements}

\begin{figure*}[tp!]
\centering
\includegraphics[width=0.66 \textwidth]{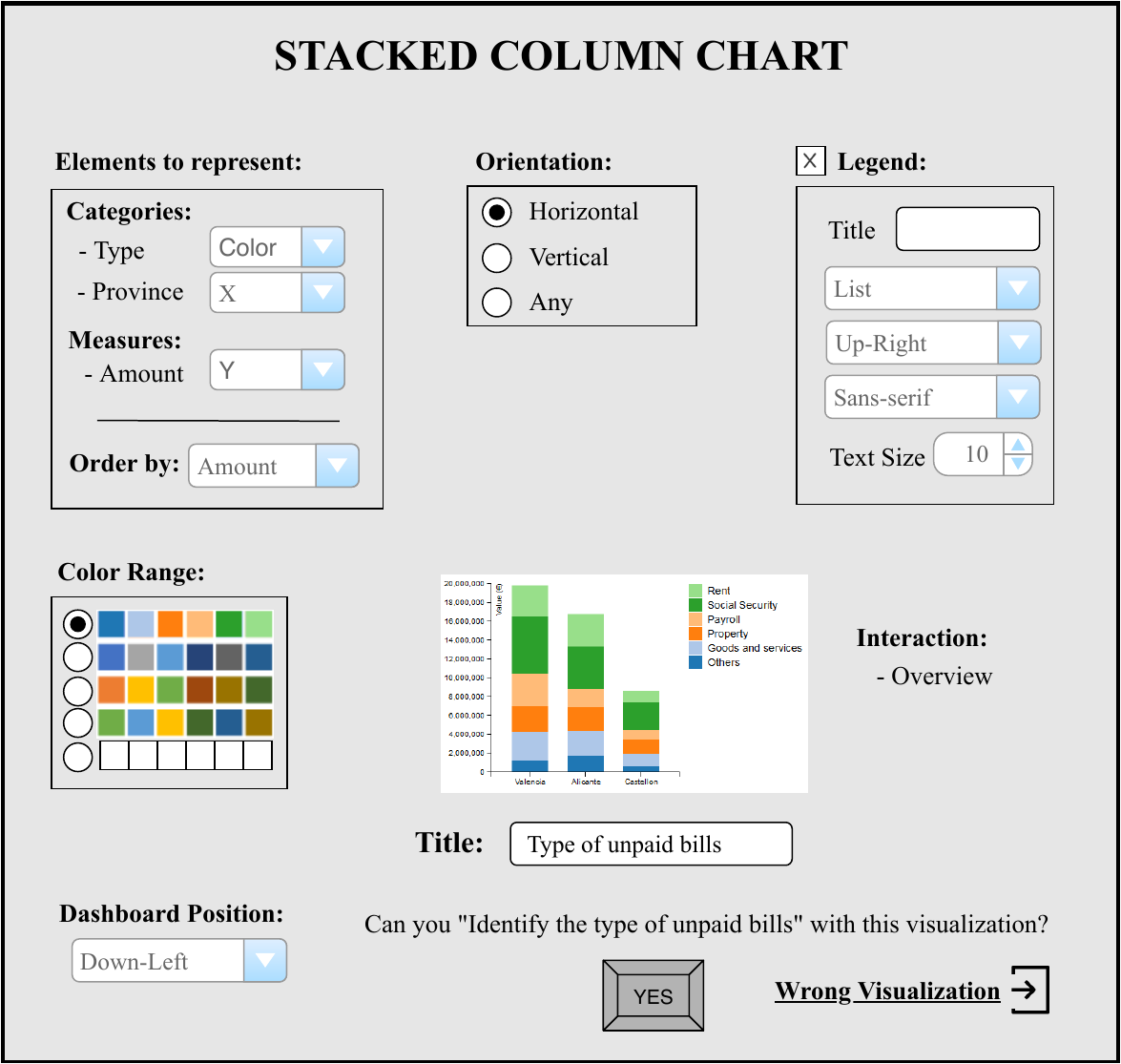}
\caption{Application of our data visualization metamodel to the case study.}
\label{fig:exprototipo}
\end{figure*}

The visualization specification is used as input of the Data Visualization Model. A visualization tool will be generated as Fig. \ref{fig:exprototipo} shows using the information collected in the process.

The tool shows the most suitable visualization type, the integration type defined by the user and a representation of the visualization. It also shows the selected elements to be represented in the visualization. The user will have to choose in which \textbf{axes} she want to see each element represented. In this case, we have \textit{``Province''} in X axis, \textit{``Amount''} in Y axis and \textit{``Type''} as Color. The user also has to select the element that determines the \textbf{order} in the visualization. Other element to specify is the \textbf{orientation} of the visualization, this can be defined as horizontal, vertical or any if the visualizations have no orientation. In this case the user has decide to user a horizontal orientation. Next element is the \textbf{legend}, which can be shown or not. A legend may have a title, a type (in this case the user has decide to represent it like a list), a position on the visualization, a font family, and a text size. The \textbf{range of colours} used to represent the visualization also has to be choose, the user can choose one of the color ranges proposed or personalize a range. Finally, the user can give a dashboard position to the visualization and a title. 

\begin{figure*}[tp!]
\centering
\includegraphics[width=0.83 \textwidth]{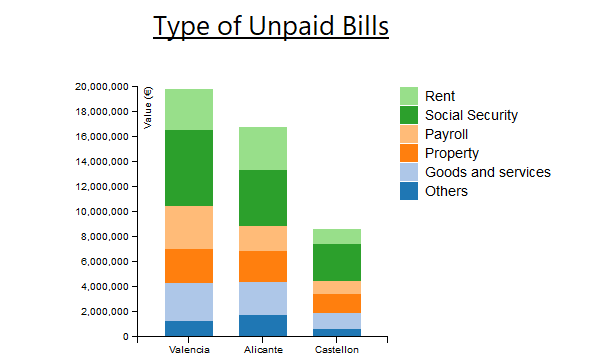}
\caption{Visualization rendered in D3.js.}
\label{fig:barras}
\end{figure*}

The user will review the data visualization model until she achieves her visualization requirements. Once all the elements have been customized, the user has to validate if the visualization obtained does contribute to answer her informational goal, in this case \textit{``Identify the type of unpaid bills''}. If the visualization is validated, it will be generated making a call to the D3 JavaScript library \cite{D3js}, obtaining the visualization shown in Fig. \ref{fig:barras}. Otherwise, an unsuccessful validation would generated a review of the existing user requirements model, to start a new iteration and generating in turn an updated model.

This visualization, combined with those generated for the others information goals, will be grouped into a dashboard, aimed at satisfying the analytic requirements of our tax collector user with the most adequate visualizations and covering all the data required by the analysis. 

\section{Conclusions and Future Work}

In this paper, we have presented an approach in the context of the Model Driven Architecture (MDA) standard in order to help users derive the most adequate visualizations. Our approach envisages three different models, i) a requirements model based on goal-oriented modeling for representing information requirements; ii) a data profiling model that abstracts the required information from the data sources; and, iii) a visualization model for capturing visualization details regardless of their implementation technology. Together with these models, we have proposed a series of transformations that allow us to bridge the gap between information requirements and the actual implementation. The great advantage of our proposal is that users can focus on their information needs and obtain the visualization that is better suited for their particular case, without requiring visualization expertise. In order to check the validity of our approach, we have applied our approach to a real use case focused on a tax collection organization. The results obtained, as well as a currently ongoing family of experiments, support the approach presented.   

As part of our future work, we are working on the definition and generation of dashboards as a whole. In this way, we will simplify and reduce the resources required to obtain visual analytics, which is of special interest for small and medium companies who cannot afford hiring several analysts in order to cover data, visualization, and business expertise required for Big Data analytics. 

\section{Acknowledgment}
This work has been co-funded by the ECLIPSE-UA (RTI2018-094283-B-C32) project funded by Spanish Ministry of Science, Innovation, and Universities. Ana Lavalle holds an Industrial PhD Grant (I-PI 03-18) co-funded by the University of Alicante and the Lucentia Lab Spin-off Company.

\bibliographystyle{splncs04}
\bibliography{mybibliography}

\end{document}